\documentclass{ptapap}

\usepackage{geometry,amsmath}

\let \sun \odot

\author{Kyle C. Augustson}[CEA]
\affil[CEA]{AIM, CEA, CNRS, Universit\'{e} Paris-Saclay, Universit\'{e} Paris Diderot, Sarbonne Paris Cit\'{e}, F-91191 Gif-sur-Yvette Cedex, France}

\title{Magnetism in Massive Stars}

\begin{document}

\maketitle

\begin{abstract}
  Massive stars are the drivers of star formation and galactic dynamics due to their relatively short lives and
  explosive demises, thus impacting all of astrophysics.  Since they are so impactful on their environments, through
  their winds on the main sequence and their ultimate supernovae, it is crucial to understand how they evolve.  Recent
  photometric observations with space-based platforms such as CoRoT, K2, and now TESS have permitted access to their
  interior dynamics through asteroseismology, while ground-based spectropolarimetric measurements such as those of
  ESPaDOnS have given us a glimpse at their surface magnetic fields. The dynamics of massive stars involve a vast range
  of scales. Extant methods can either capture the long-term structural evolution or the short-term dynamics such as
  convection, magnetic dynamos, and waves due to computational costs.  Thus, many mysteries remain regarding the impact
  of such dynamics on stellar evolution, but they can have strong implications both for how they evolve and what they
  leave behind when they die. Some of these dynamics including rotation, tides, and magnetic fields have been addressed
  in recent work, which is reviewed in this paper.
\end{abstract}

\section{Introduction}

Massive stars live fast and die young. Because of this fast lifecycle, massive stars have been the primary drivers of
galactic evolution and to some degree cosmological evolution from the the epoch of reionization and the formation of the
first stars. During the main sequence, where stars spend most of their lives burning hydrogen in their cores, the strong
winds of these hot stars impact their local environment and any stellar companions that may have formed nearby.  When
these stars die in a brilliant supernova, their angular momentum, magnetic fields, and heavy-element laden ionized ashes
are redispersed into the local medium \citep[e.g.,][]{maeder09,langer12}.  This eventually leads to an enrichment of
galaxy in elemental abundances and triggers new episodes of star formation, although with a modified abundance
distribution that impacts the nature of the stars that form.  But the precise statistical behavior of this process is an
important open question in galactic dynamics and cosmological evolution \citep[e.g.,][]{notomo13,krumholz19}, one that
future studies can help to address.  Moreover, those explosive events, often leave behind a remnant: a white dwarf for
lower mass stars, a neutron star for intermediate mass stars, and black holes for the more massive stars, whereas some
of the most massive stars may completely disintegrate in a titanic pair-instability driven explosion
\citep[e.g.,][]{woosley07,groh13}.  Indeed, one of the mysteries of these objects is why only a fraction of these white
dwarf and neutron star remnants have extremal surface magnetic fields, whereas the remaining fraction have comparatively
weak magnetic fields \citep[e.g.,][]{donati09,mosta15,kaspi17}.\\

Throughout the evolution of massive stars, there are processes that can build internal magnetic fields and transport
angular momentum and chemical species \citep[e.g.,][]{maeder09,mathis13}. The rotation and magnetic fields of these
stars drastically impact both their evolution through modified convective, transport processes, and mass loss and their
environment through the strong winds associated with that mass loss and their tidal interactions with any companions
\citep[e.g.,][]{ogilvie14,smith14}.  Since these dynamical processes affect the long-term evolution of such stars, they
must be modeled with high fidelity in order to properly capture their impact on many other astrophysical processes.
Observationally calibrating these models is possible given the recent revolution in our knowledge of stellar dynamics
provided by the seismology of the interiors of the Sun and of stars (with SDO, CoRoT, Kepler, K2, TESS, and BRITE) and
through the ground-based spectropolarimetry that characterizes stellar surface magnetic fields (ESPaDOnS/CFHT,
Narval/TBL, HARPSpol/ESO).\\

\begin{figure}[t!]
  \begin{center}
    \includegraphics[width=0.95\textwidth]{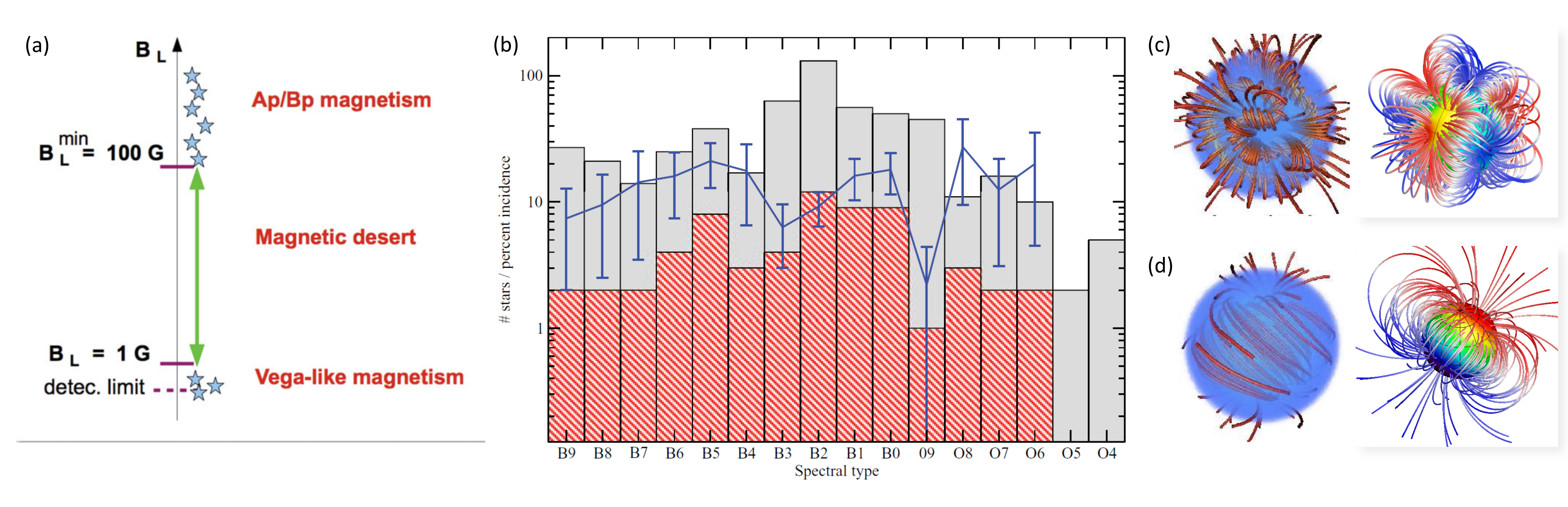}
    \caption{(a) The magnetic desert, no detections of stars with magnetic fields between about 1G and 100 G
      \citep{lignieres14}. (b) Distribution of stars with detected magnetic fields by stellar type \citep{wade14}. (c)
      Simulations of relaxed magnetic fields for a nearly uniform initial magnetic field amplitude \citep{braithwaite08}
      and a similar observation \citep{kochukhov11}. (d) Similar to (c) but for a more centrally concentrated initial
      condition, with corresponding observation \citep{kochukhov13}.}\label{fig:mag}
    \vspace{-0.5cm}
 \end{center}
\end{figure}

\noindent\textit{Magnetism}\\
Spectropolarimetric campaigns by consortia such as MiMeS (Magnetism in Massive Stars) and BOB (B Fields in OB Stars) and
LIFE (Large Impact of magnetic Fields on the Evolution of hot stars) have been directed toward measuring magnetic fields
on the surfaces of massive stars, some using Zeeman Doppler imaging techniques. They report that only about 7\% of O and
B-type stars exhibit large-scale surface magnetic fields \citep[See Figure \ref{fig:mag}; e.g.,][]{donati09,
  wade14,fossati15,fossati16}. As for interior magnetic fields, strong magnetic fields have potentially been detected
deep in the cores of these stars through the suppression of dipolar mixed oscillatory modes
\citep{fuller15,stello16}. Such depressed modes are seen in those stars that had a convective core during the main
sequence, suggesting that they were indeed running a convective core dynamo.  Additionally, an asteroseismic method for
detecting general magnetic field configurations in the interiors of rapidly rotating stars has been developed for
gravity waves and applied to Kepler stars to ascertain if the frequency shifts can be detected \citep{prat19}, and for
perturbative rotation and magnetic fields for general stellar waves in \citet{augustson18}.  Hence, the tools are in
hand to assess data from ongoing and upcoming ground-based and space-based observational campaigns for the internal
structure and magnetic fields of stars. However, both the processes that lead to strong angular momentum transport and
to convective dynamos remain difficult to universally parameterize so as to explain the observed properties and the
secular evolution of massive stars.

\begin{figure}[t!]
  \begin{center}
    \includegraphics[width=0.95\textwidth]{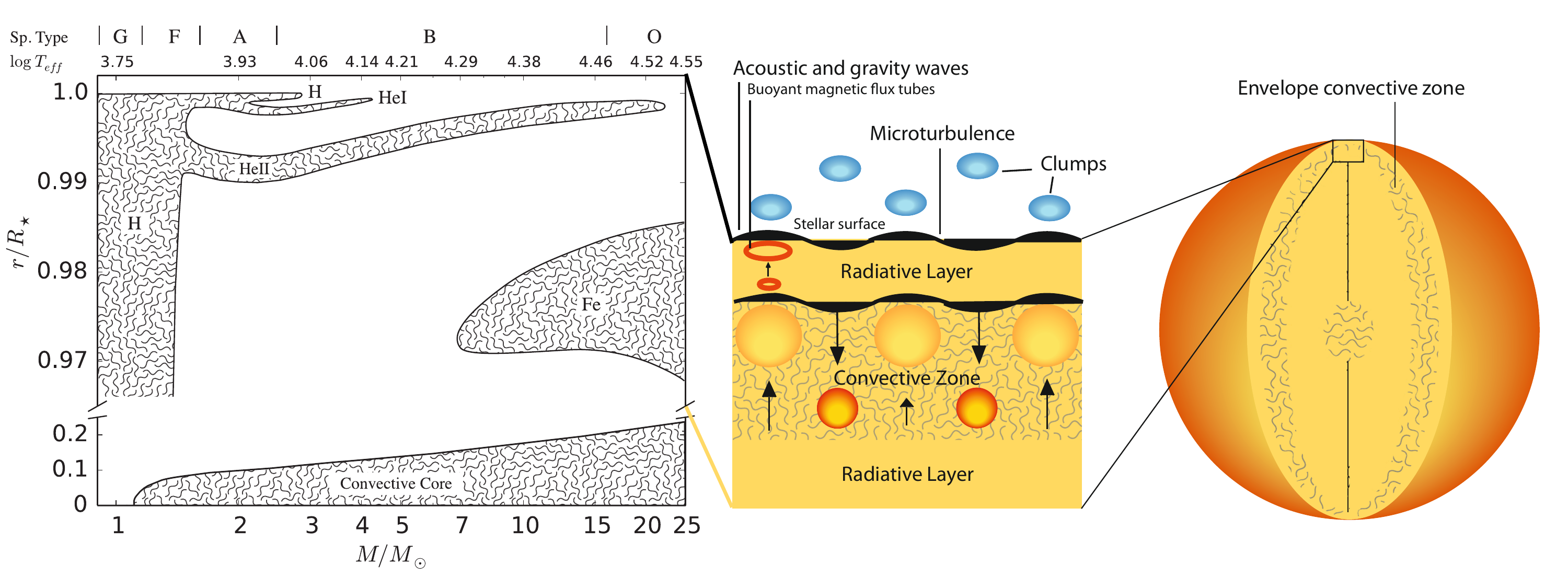}
    \caption{Left: Normalized radial extent of the core, surface, and subsurface convective zones for stars between 0.9 and
      $25 M_{\sun}$, corresponding to the main sequence structure of these stars.  The convective regions are associated
      with the ionization of H, HeI, HeII, and the iron group elements Fe.  The stellar surface $r/R_*$ is defined as
      the location where the optical depth $\tau = 2/3$ \citep{cantiello19}. Right: A sketch of the iron-bump convection
      zone, showing a local box in the spherical domain with a portion of the inner radiative zone, FeCZ, and outer radiative
      photosphere where waves are excited and observed as macroturbulence \citep{cantiello09}.}\label{fig:fecz}
    \vspace{-0.5cm}
 \end{center}
\end{figure}

\noindent\textit{Convection}\\
Throughout the evolution of massive stars, convection has profound effects on both their stellar structure and on what
we observe at the surface. In main-sequence massive stars, the photosphere is in a stably-stratified region where
radiation dominates the heat transport and convective motions are absent. Yet observations show significant motions of
unknown origin called macroturbulence at the stellar photosphere, with typically supersonic velocities of
$20-60~\mathrm{km~s^{-1}}$ \citep{sundqvist13}. A possible source of it may be a detached convection zone located well
below the photosphere, where iron has a local maximum in its opacity
\citep[e.g.,][]{cantiello11,cantiello19,nagayama19}. These iron-bump convection zones host nearly sonic compressive
convection, driving waves in the overlying region (see Figure \ref{fig:fecz}). However, there is a curious targe called
``Dash-2.''  This star is an outlier among the observed massive stars with an observed macroturbulence of only
$2.2~\mathrm{km~s^{-1}}$. It also has the strongest observed surface magnetic field of these stars, with a surface field
of approximately 20~kG, versus typical field strengths of 1~kG \citep{sundqvist13}. Dash-2’s surface magnetic fields are
strong enough that they rival the thermodynamic pressure in the iron-bump convection zone, potentially quenching the
convection and thus the surface waves in Dash-2. In the other stars of \citet{sundqvist13}, the fields are too weak
relative to the thermodynamic pressure. Hence, one puzzle to solve is the origin of macroturbulence and its link to
near-surface convection and the influence of magnetism.  The iron-bump convection zone is however only one of the
convective regions, the deeper convection zone and the seat of a global-scale dynamo is the convective core.  This core
convection will drive internal waves and interact with the fossil magnetic field
\citep[e.g.,][]{featherstone09,augustson16}.  Such dynamo action will have impacts later in the evolution of the star as
the internal structure of the star freezes the dynamo-generated field into a larger stable
region, adding to the extant fossil field there.\\

\begin{figure}[t!]
  \begin{center}
    \includegraphics[width=0.95\textwidth]{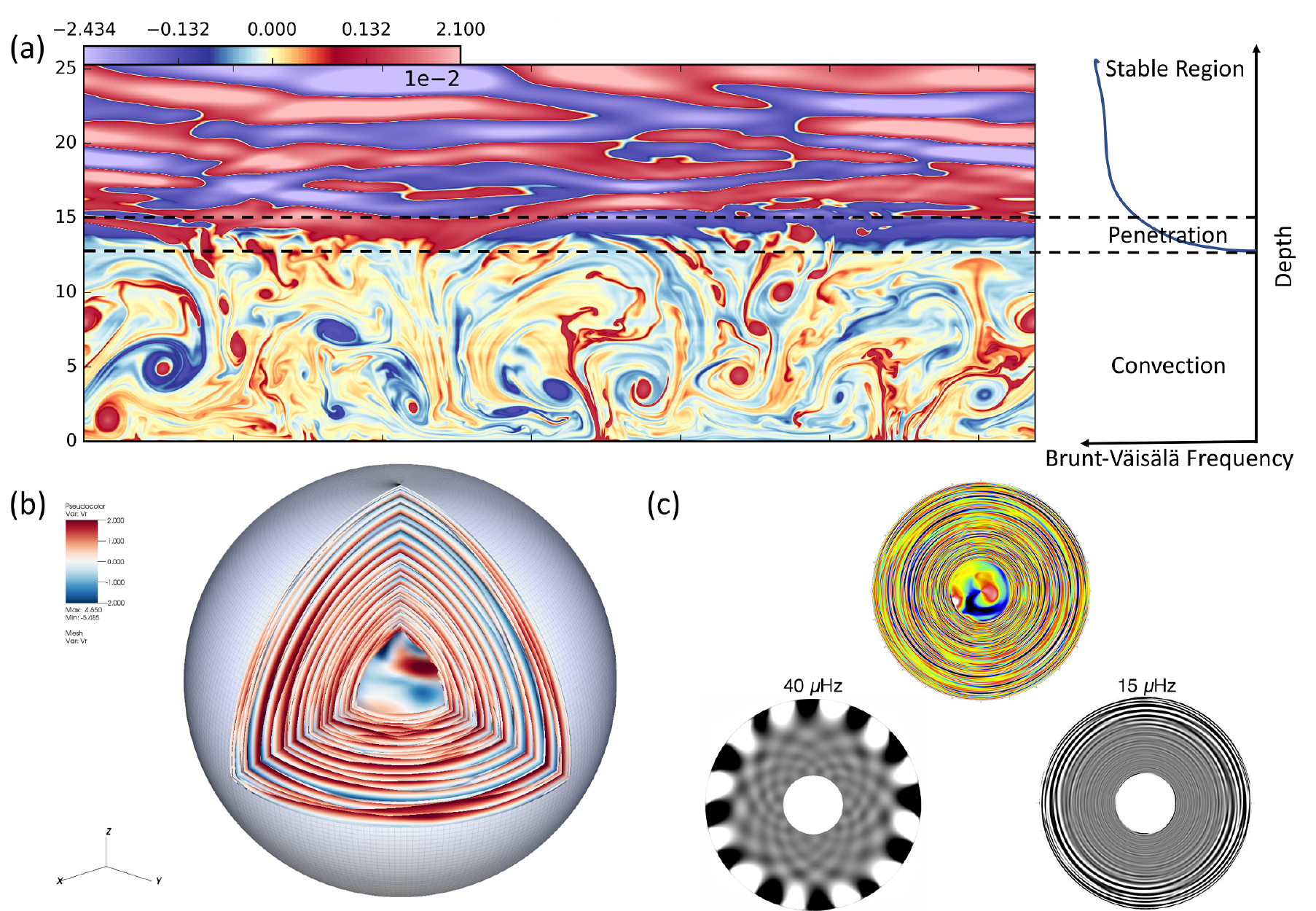}
    \caption{(a) A Dedalus simulation of fully compressive convection below a stable region, reminiscent of both a
      convective core and the iron-bump convection zone, showing the convective region, penetration into the stable
      region, and the excited waves in the stable region as rendered in entropy fluctuations with cool regions in dark
      tones and warm regions in light tones (Courtesy of B. Brown). (b) Radial velocity of waves excited by a convective
      core in a 3D simulation of a $15~M_{\sun}$ star, showing a displacement proportional to the wave amplitude
      \citep{andre19}. (c) An equatorial cut through the computational domain of the $15~M_{\sun}$ star, showing the
      filtered extraction of waves at two frequencies \citep{andre19}.}\label{fig:pene}
    \vspace{-0.5cm}
 \end{center}
\end{figure}

\noindent\textit{Convective Penetration}\\
Convective flows cause mixing not only in regions of superadiabatic temperature gradients but in neighboring
subadiabatic regions as well (see Figure \ref{fig:pene}(a)), as motions from the convective region contain sufficient
inertia to extend into those regions before being buoyantly braked or turbulently eroded
\citep[e.g.,][]{augustson19a,korre19}. Thus, convective penetration and turbulence softens the transition between
convectively stable and unstable regions, with the consequence being that the differential rotation, opacity,
compositional and thermodynamic gradients are modified \citep[e.g.,][]{augustson13,brun17,pratt17}, while compositional
gradients can drive further mixing \citep[e.g.,][]{garaud18,sengupta18}.  Such processes have an asteroseismic signature
as has been observed in massive stars as well as lower mass stars
\citep[e.g.,][]{aerts03,neiner12,neiner13,moravveji16,pedersen18}. Indeed, the waves shown in Figure \ref{fig:pene}(b)
and (c) depict the self-consistent amplitude of waves excited by convection in the core. In stars with a convective
core, convective penetration can lead to a greater amount of time spent on stable burning phases as fresh fuel is mixed
into the burning region \citep[e.g.,][]{maeder09,viallet13,jin15}.  From the standpoint of stellar evolution, this is
yet an open problem: to understand how the depth of penetration and the character of the convection
in this region change with rotation, magnetism, and diffusion.\\

\begin{figure}[t!]
  \begin{center}
    \includegraphics[width=0.95\textwidth]{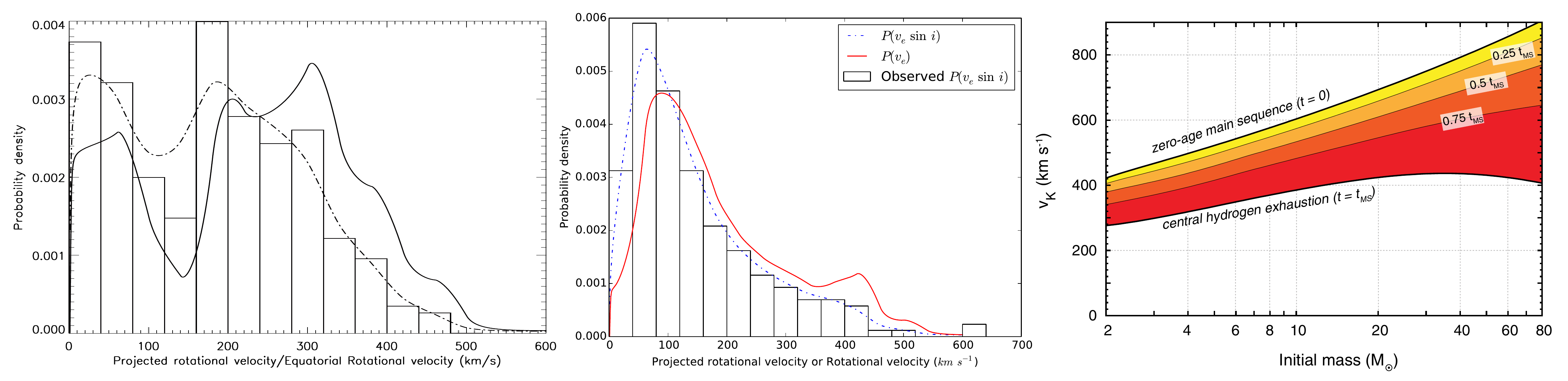}
    \caption{Left: Observed bimodal distribution of $v\sin{i}$ for single B-type stars \citep{dufton13}. Center:
      Observed log-normal distribution of $v\sin{i}$ for single O-type stars \citep{ramirez13}. Right: Theoretical
      Keplerian velocities of massive stars as a function of age and mass \citep{demink13}.}\label{fig:rot}
    \vspace{-0.5cm}
 \end{center}
\end{figure}

\noindent\textit{Rotation}\\
Observations using stellar spectra have shown that the majority of massive stars are fairly rapid rotators
\citep[e.g.,][]{huang10,demink13}. The average projected equatorial velocity of these stars is about
$150~\mathrm{km~s}^{-1}$ on the main-sequence but have significant tails (Figure \ref{fig:rot}). The rotation rates of
the core and radiative envelope of some B stars, notably $\beta$-Cepheid variables, have been estimated using
asteroseismology, finding that most massive main sequence stars appear to be rigidly rotating \citep[e.g.,][]{aerts17}.
Indeed, the small angular velocity contrasts for observed massive stars may point to an effective angular momentum
coupling between the core and the envelope, possibly by gravity waves or magnetic fields during their evolution.
Stellar radiative regions are rotating and magnetized. Therefore, internal gravity waves become magneto-
gravito-inertial waves and the Coriolis acceleration and the Lorentz force cannot be treated apriori as
perturbations. For example, during the PMS of low-mass stars and in rapidly rotating massive stars, the stratification
restoring force and the Coriolis acceleration can be of the same order of magnitude. In addition to the impact of
rotation on convection and its dynamo processes, it also has an influence in radiative zones that can lead to transport
there through meridional flows, shear turbulence, and internal waves \citep{mathis13}. The large-scale meridional
circulation occurring in stellar radiation zones is occurs due to the deformation of the star and its isothermal
surfaces by the centrifugal acceleration. The radiative flux is then no longer divergence-free and must be balanced by
heat advection, which is carried by the meridional flow.  This flow can also transport angular momentum and chemical
species throughout the radiative envelope.  Shear turbulence can occur if the waves excited by convection become
nonlinear and break or if there is a strong differential rotation that leads either to an magneto-rotational
instability.  Such turbulence has been successful in describing some aspects of the dynamics in massive stars
\citep{meynet00}. Waves also can transport energy, even if they are linear as they can propagate until they are
dissipated through thermal diffusion. Thus, their transport is highly nonlocal. Such internal waves are excited by the
turbulent motions at convection to radiation transitions in stellar interiors, namely the boundaries of convective
envelopes.\\

\begin{figure}[t!]
  \begin{center}
    \includegraphics[width=0.95\textwidth]{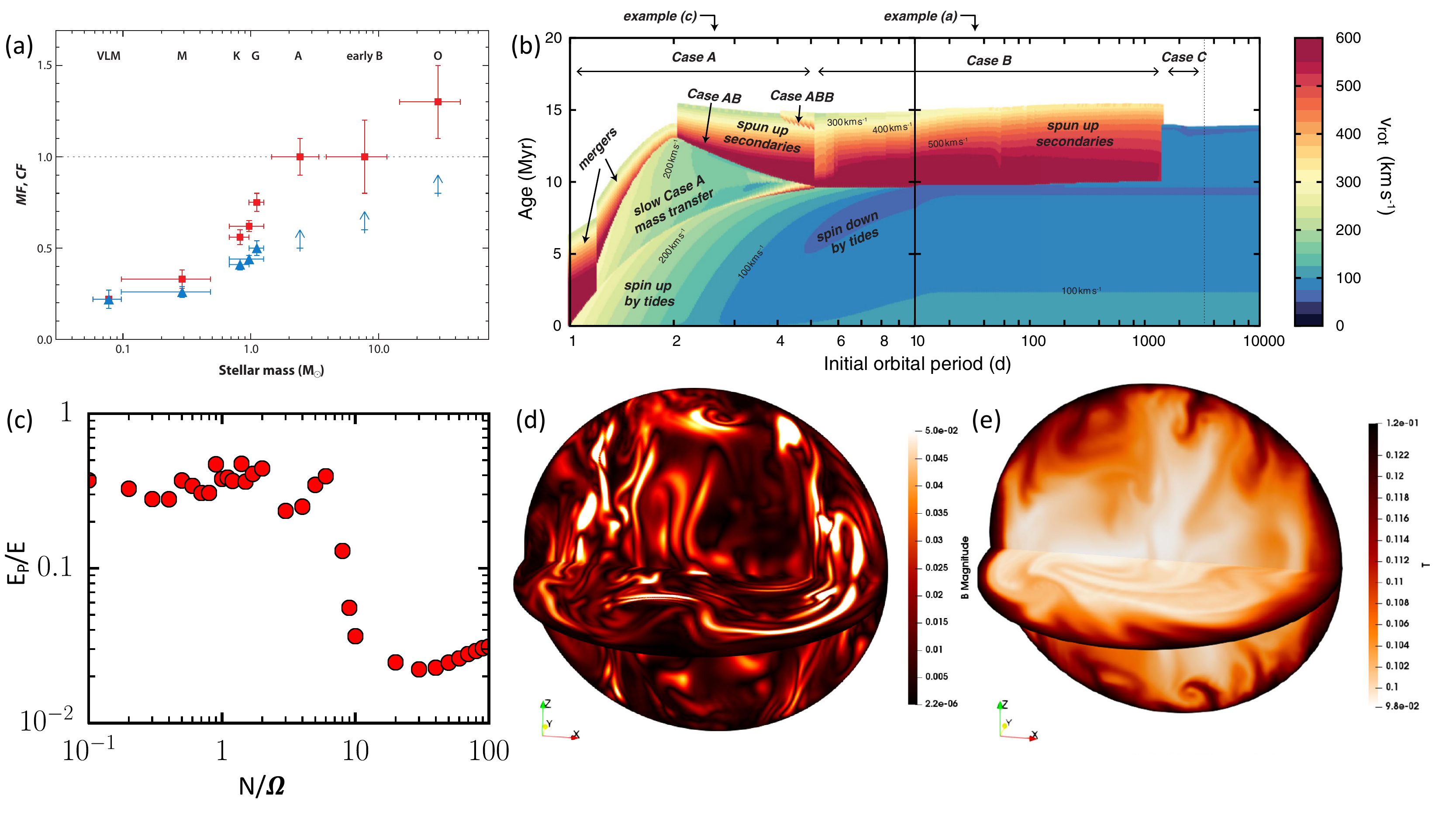}
    \caption{(a) Average companion fraction (CF) and multiplicity fraction (MF) by spectral type \citep{duchene13}. (b)
      Rotational evolution and outcomes of a massive binary system as function of initial orbital period
      \citep{demink13}. (c) Ratio of poloidal magnetic energy to total energy in simulations of tidally driven dynamos
      in a stable region, (d) magnetic field strength for $N/\Omega=0.5$ and (e) its corresponding temperature field
      \citep{vidal18}.}\label{fig:tides}
 \end{center}
\end{figure}

\noindent\textit{Multiplicity}\\
Recent surveys have shown that around 70-80\% of massive stars are in multiple star systems
\citep{raghavan10,duchene13}. Given that their lives are short, they do not have much time to migrate far from their
place of birth. Therefore, tidal interactions will be an important dynamical component for many massive stars, which
will impact their evolution, structure, and magnetic fields. The equilibrium tides distort the shape of the star while
it is the dynamical tides, or even nonlinear tides, that could lead to dynamo action if there is sufficient correlation
between the velocity field and the magnetic field \citep[See Figure \ref{fig:tides};][]{ogilvie14,vidal18,vidal19}. Such
processes therefore should be accounted for in both 3D dynamical simulations of massive stars as well as in stellar
evolution and structure computations. Indeed, as shown in Figure \ref{fig:tides}(c), the poloidal magnetic energy
generated through dynamo action induced by the tide can reach 20\% of the equipartition value with the kinetic energy
of the dynamical tide when the orbital frequency greater than about 10\% of the Brunt-V\"{a}is\"{a}ll\"{a} frequency.
This suggests that it is possible that tides could disrupt the fossil fields formed during the pre-main-sequence.
Moreover, it implies that the tidal interaction between the disk and the protostar is indeed quite dynamic even in the
stably stratified regions of the protostar. \\

\section{Evolution of Massive Star Magnetism}

\begin{figure}[t!]
  \begin{center}
    \includegraphics[width=0.95\textwidth]{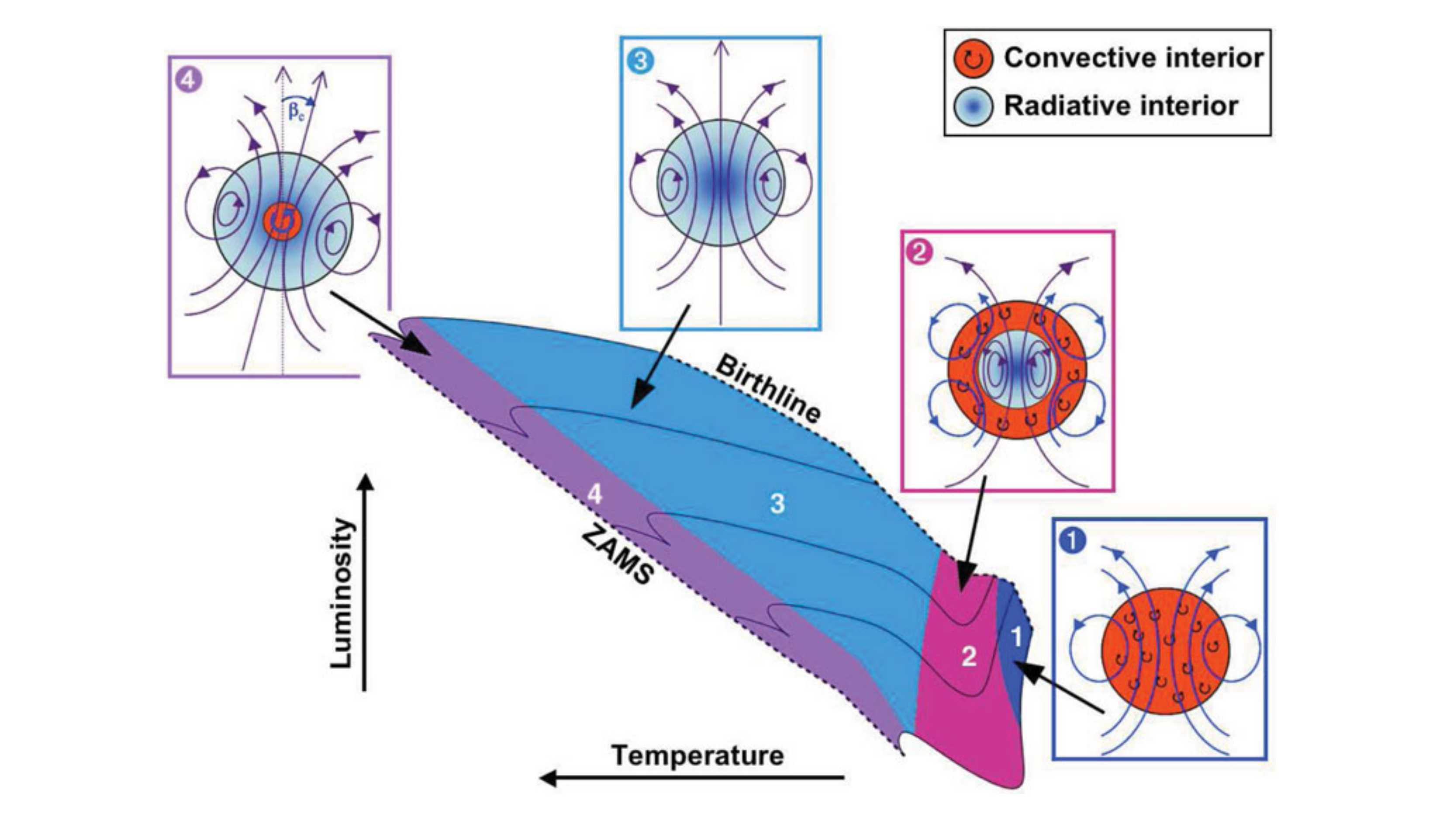}
    \caption{A sketch of the evolution of a PMS massive star, showing the fully convective phase (1), the convective
      freeze-out (2), the radiative phase (3), and the ZAMS state with an oblique axisymmetric relaxed magnetic field
      (4) \citep{neiner15}.}\label{fig:pms}
    \vspace{-0.5cm}
 \end{center}
\end{figure}

The formation of massive stars is quite different from the slow accretion of low mass stars that can take 100 million
years, with the formation time scale being compressed to a few tens or hundreds of thousands of years. What they do
share is that whatever angular momentum, chemical abundances, and magnetic field are present in the star forming region
will initially shape the structure of the local patch of gas that if sufficiently self-gravitating will eventually
collapse into a disk that feeds a central object. As gravitational energy is released while this central object
contracts and it is fed with additional mass from the disk, parts of the protostar will be convectively unstable plasma
(see Figure \ref{fig:pms}).

The stably stratified portions of the protostar will contain the geometrically amplified magnetic field advected into it
during its initial condensation.  The convective portions on the other hand will be running an active magnetic dynamo
wherein the rotation, differential rotation, and the buoyantly driven convection act together to build magnetic fields
that can be stronger than the originating field.  The dynamo-generated field will link with the magnetic field in the
stable regions of the star as well as with the magnetic fields in the disk causing angular momentum transfer as well as
potentially inciting instabilities \citep[e.g.,][]{romanova15}.  This formation process is extremely difficult both to observe
and to theoretically describe given that the structure of the disk impacts the protostellar structure so strongly,
e.g. it is unknown how the mass, angular momentum, and magnetic field are actually entrained into the protostar.

Massive protostars begin fusing material in their cores before they finish forming leading to an even larger radiative
flux compared to their low-mass brethren that radiate only the gravitational energy. The radiation streaming from the
massive protostar's photosphere is sufficient to blow away most of the material unless the disk is some how screened
from it. One way this can be circumvented is through the magnetic collimation of polar jets of outflowing gas and dust
along which the radiation can preferentially stream. These jets can drive a circulation in the disk that draws in more
gas from its surroundings \citep[e.g.,][]{tan14,krumholz15,romanova15}, replenishing the disk that feeds the star. This
process continues adding mass and angular momentum to the star until its radiation and wind output increases enough to
overpower mass inflowing from the disk, stalling that flow and eventually eviscerating the disk.

\begin{figure}[t!]
  \begin{center}
    \includegraphics[width=0.95\textwidth]{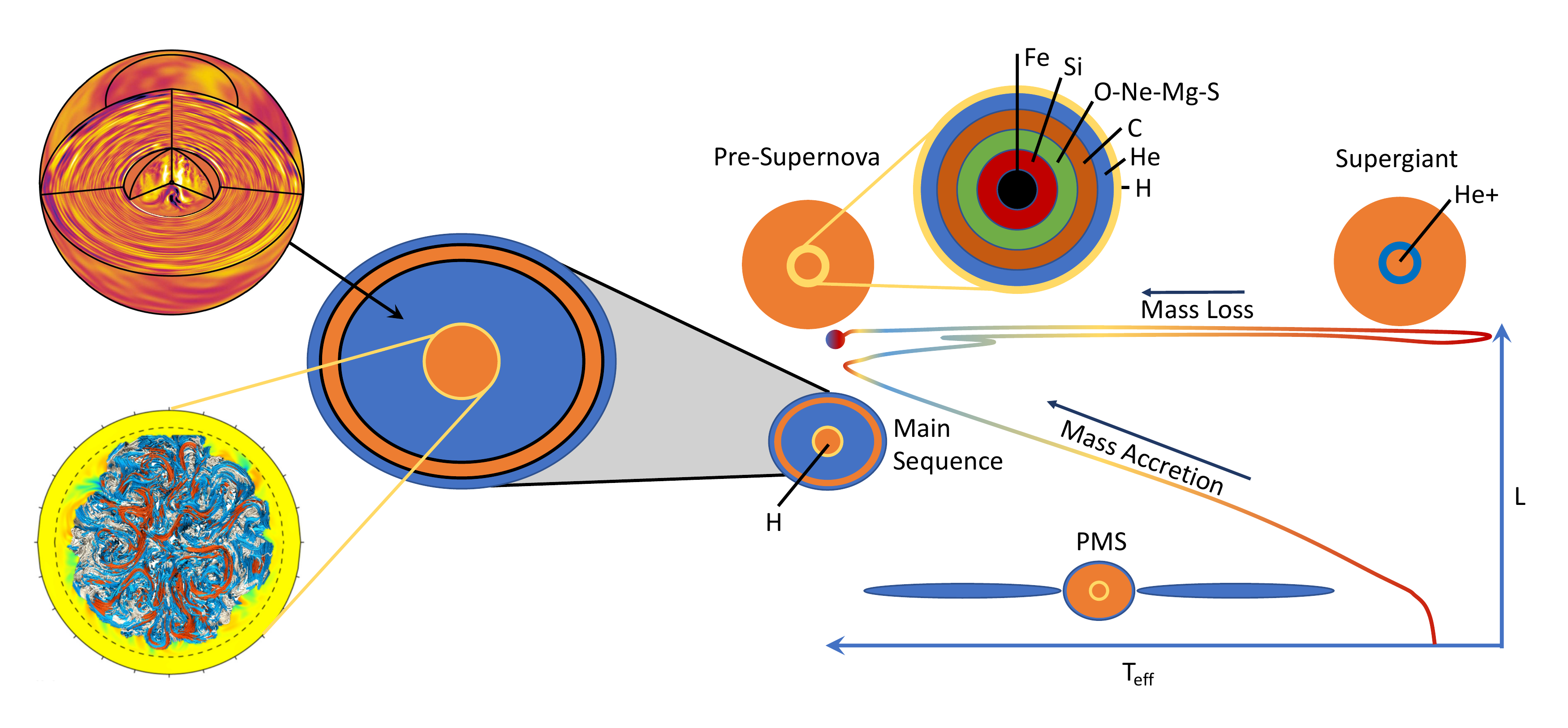}
    \caption{A sketch of the evolution of a massive star, showing 3D convection and its associated transport and dynamo
      processes and the inclusion of 3D processes occurring in radiative regions being linked into the parameterized
      processes of main sequence evolution of a 2D star.  The 3D processes are depicted here as magnetic field lines in
      the equatorial plane with strength indicated as 1 G (gray), 100 kG (blue), and 1 MG (red) for Task A, and as
      normalized radial velocity, showing the wave field excited in the radiative envelope by the convective core with
      downflows in dark tones and upflows in light tones \citep{augustson16}. The path on the left of the diagram
      illustrates the evolution of a rotating $40M_{\sun}$ star, from the accreting pre-main-sequence phase, to the
      main-sequence, to the post-main-sequence, and ultimately to its supernova.}\label{fig:fig1}
    \vspace{-0.5cm}
 \end{center}
\end{figure}

Once the massive protostar has fully contracted, disconnected from the disk, and begun its CNO-cycle fusing
main-sequence life, its magnetic history is locked into its convectively stable regions as a fossil field that is
connected to the convective dynamo of its core and into its radiation driven winds.  The topological properties of this
fossil field can impact whether or not the star has a magnetosphere.  This in turn affects the star's mass loss rates
and the rate of its spin down. Moreover it can impact the nature of its near-surface convection zones and the waves it
generates that manifest as macroturbulence \citep{sundqvist13,macdonald19}.  Thus, understanding the properties of this
fossil field is paramount.

\subsection{Fossil Magnetic Fields}

One current puzzle regarding massive star magnetism is why do only about ten percent of such stars possess observable
magnetic fields.  Could it be that 90 percent have complex morphologies that do not lend themselves to
spectropolarimetric detection?  Or is it that there are configurational instabilities that lead to only a subset of
stable magnetic configurations? The stability of magnetic field configurations for certain stratified fluid domains have
been considered both from a theoretical and a numerical standpoint in the work of \citet{duez10a} and \citet{braithwaite04,duez10b},
respectively.  These magnetic field equilibria are valid for a spherically-symmetric barotropic star, with the
nonbarotropic component being handled perturbatively assuming that the magnetic field is such that the magnetic pressure
is much less than the gas pressure everywhere in the domain considered.  Such magnetic equilibria are shown in Figure
\ref{fig:mag}(c) and (d).

One crucial physical component that has been neglected so far is rotation for most ZAMS massive stars will be rapidly
rotating.  Thus, if one considers rotation as discussed in \citet{duez11} and \citet{emeriau15}, and with the
details forthcoming in \citet{mathis20}, there are additional ideal invariants in the system that permit the
construction of self-consistent magnetic and mean velocity equilibria in the rotating frame. The underlying principle
for finding magnetic equilibria in ideal magnetohydrodynamics is that the energy and dynamics must be fixed in time. For
this to be true, the Lorentz force must be in balance with the other forces in the system: the pressure, gravity, and
the Coriolis force, forming a magneto-rotational hydrodynamic equilibrium.  Numerical simulations of such equilibria in
rotating systems appear to yield relaxed states similar to those in the nonrotating system, but where slowly rotating
systems tend toward misalignment of the magnetic field with respect to the rotation axis and more rapidly rotating
systems are aligned \citep{duez11}. However, the stability of such systems appears to be quite sensitive to their
initial distributions of magnetic helicity, which has already been seen in numerical simulations of magnetic field
relaxation \citep[e.g.,][]{braithwaite08,braithwaite17}.

\begin{figure}[!t]
  \centering 
  \includegraphics[width=0.95\linewidth]{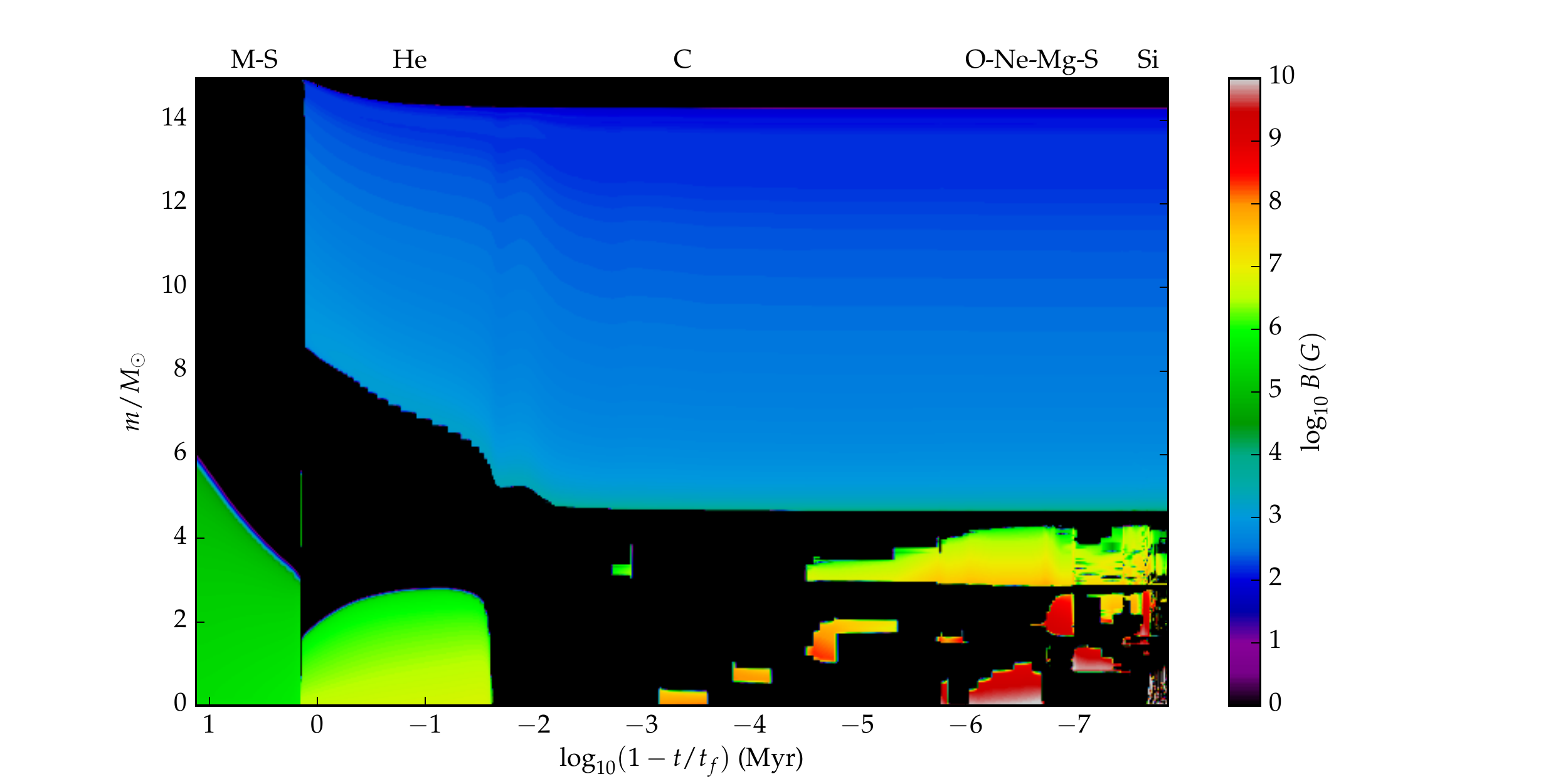}
  \caption{A magnetic Kippenhahn diagram showing the evolution of the equipartition magnetic field for a 15~$M_{\odot}$
    star. The abscissa show the time remaining in Myr before the iron core infall that occurs at $t_f$. The burning
    phase of the core is indicated at the top of the diagram.}
  \label{fig:kipp}
  \vspace{-0.5cm}
\end{figure}

\subsection{Dynamo-generated Magnetic Fields}

In convectively unstable regions, the buoyancy-driven plasma motions give rise to magnetic induction through turbulent
correlations that can generate both large and small scale magnetic fields.  Such fields can link to the fossil fields
threading through such regions.  In simulations of core convection, it has been found that superequipartition states can
be achieved where the magnetic energy is greater than that contained in the convection itself, when the rate of rotation
is sufficiently large and a the system enters into a magnetostrophic regime \citep{augustson16}. The structure of the
magnetic field is shown in an equatorial slice of the convective core on the left hand side of Figure
\ref{fig:fig1}. Such states are possible because the Lorentz force that could otherwise quench the flows is mitigated by
the flow and magnetic structures that are formed.  Specifically, the flow and magnetic structures are largely spatially
separated in that the bulk of the magnetic energy exists in regions where there is little kinetic energy and visa versa
\citep[e.g.,][]{featherstone09,augustson16}.  In the regions where the two fields overlap, new magnetic field is
generated through induction.  The presence of a fossil field can modify this balance, enhancing certain correlations and
leading to a greater level of superequipartition \citep{featherstone09}.  Such magnetic fields formed during the main
sequence will slowly be added to the fossil magnetic field in the radiative envelope of these massive stars as the
convective core contracts.  This process is continuous but occurs over evolutionary time scales, whereas the new
magnetic field configuration will relax on Alfv\'{e}nic time scales.  Such continuous magnetic field relaxation could be at
the origin of the declining prevalence of magnetic fields during the main sequence evolution as shown in
\citep{fossati16}.

As the star evolves past the main sequence, it moves directly to helium burning once the hydrogen has been exhausted in
the core.  During this phase, the core contracts further and the density is larger, leading to greater kinetic energy in
the convective core.  The superequipartition magnetic fields now approximately contain approximately 100 times the
energy compared to the main sequence.  As the evolution continues, and carbon burning takes place, the density is once
again very much larger and the superequipartition magnetic fields will contain again 100 times the energy of the helium
burning phase.  This pattern continues until the end-stage silicon burning, where the magnetic fields are of the order
of $10^{10}$ Gauss or more depending upon the previous stages of burning.  This evolutionary progression is shown in
Figure \ref{fig:kipp} for a 15 $M_{\sun}$ star for magnetic fields that are simply equipartition.

\section{Conclusions}

There are still many puzzles to solve regarding the influence of magnetic fields on the evolution and structure of
massive stars as well as on their winds and environment.  Nevertheless, as the numerous physical mechanisms at work are
explored, a picture can be constructed about both their births as well as their evolution toward their cataclysmic ends.
Here we have explored recent work regarding the observed properties of the magnetism of main-sequence massive stars, the
influence of convection on their surface properties as well as on their core dynamo action, the impact of convective
penetration both on wave generation and in chemical mixing, and finally how tides from multiple stellar companions can
change the classical picture of single star evolution. Later we have sketched the magnetic evolution of the interior of
the a massive star from the pre-main-sequence to the final stages before its supernova.

\acknowledgements{K.~C. Augustson acknowledges support from the ERC SPIRE 647383 grant.}

\bibliographystyle{ptapap}
\bibliography{augustson_jdl}

\begin{thebibliography}{67}
\providecommand{\natexlab}[1]{#1}
\providecommand{\url}[1]{\texttt{#1}}
\providecommand{\urlprefix}{URL }
\providecommand{\eprint}[2][]{\url{#2}}

\bibitem[{{Aerts} et~al.(2017){Aerts}, {Van Reeth}, \& {Tkachenko}}]{aerts17}
{Aerts}, C., {Van Reeth}, T., {Tkachenko}, A., \emph{\apjl} \textbf{847}, 1, L7
  (2017)

\bibitem[{{Aerts} et~al.(2003)}]{aerts03}
{Aerts}, C., et~al., \emph{Science} \textbf{300}, 5627, 1926 (2003)

\bibitem[{{Andr\'{e}}(2019)}]{andre19}
{Andr\'{e}}, Q., {Ondes gravito-inertielles dans les \'{e}toiles et les
  plan\`{e}tes g\'{e}antes}, Ph.D. thesis, l'Universit\'{e} de Paris (2019)

\bibitem[{{Augustson} et~al.(2013){Augustson}, {Brun}, \&
  {Toomre}}]{augustson13}
{Augustson}, K.~C., {Brun}, A.~S., {Toomre}, J., \emph{\apj} \textbf{777}, 2,
  153 (2013)

\bibitem[{{Augustson} et~al.(2016){Augustson}, {Brun}, \&
  {Toomre}}]{augustson16}
{Augustson}, K.~C., {Brun}, A.~S., {Toomre}, J., \emph{\apj} \textbf{829}, 2,
  92 (2016)

\bibitem[{{Augustson} \& {Mathis}(2018)}]{augustson18}
{Augustson}, K.~C., {Mathis}, S., in SF2A-2018: Proceedings of the Annual
  meeting of the French Society of Astronomy and Astrophysics, Di (2018)

\bibitem[{{Augustson} \& {Mathis}(2019)}]{augustson19a}
{Augustson}, K.~C., {Mathis}, S., \emph{\apj} \textbf{874}, 1, 83 (2019)

\bibitem[{{Braithwaite}(2008)}]{braithwaite08}
{Braithwaite}, J., \emph{\mnras} \textbf{386}, 4, 1947 (2008)

\bibitem[{{Braithwaite} \& {Spruit}(2004)}]{braithwaite04}
{Braithwaite}, J., {Spruit}, H.~C., \emph{\nat} \textbf{431}, 7010, 819 (2004)

\bibitem[{{Braithwaite} \& {Spruit}(2017)}]{braithwaite17}
{Braithwaite}, J., {Spruit}, H.~C., \emph{Royal Society Open Science}
  \textbf{4}, 2, 160271 (2017)

\bibitem[{{Brun} et~al.(2017)}]{brun17}
{Brun}, A.~S., et~al., \emph{\apj} \textbf{836}, 192 (2017)

\bibitem[{{Cantiello} \& {Braithwaite}(2011)}]{cantiello11}
{Cantiello}, M., {Braithwaite}, J., \emph{\aap} \textbf{534}, A140 (2011)

\bibitem[{{Cantiello} \& {Braithwaite}(2019)}]{cantiello19}
{Cantiello}, M., {Braithwaite}, J., \emph{\apj} \textbf{883}, 1, 106 (2019)

\bibitem[{{Cantiello} et~al.(2009)}]{cantiello09}
{Cantiello}, M., et~al., \emph{\aap} \textbf{499}, 1, 279 (2009)

\bibitem[{{de Mink} et~al.(2013)}]{demink13}
{de Mink}, S.~E., et~al., \emph{\apj} \textbf{764}, 2, 166 (2013)

\bibitem[{{Donati} \& {Landstreet}(2009)}]{donati09}
{Donati}, J.~F., {Landstreet}, J.~D., \emph{\araa} \textbf{47}, 1, 333 (2009)

\bibitem[{{Duch{\^e}ne} \& {Kraus}(2013)}]{duchene13}
{Duch{\^e}ne}, G., {Kraus}, A., \emph{\araa} \textbf{51}, 1, 269 (2013)

\bibitem[{{Duez}(2011)}]{duez11}
{Duez}, V., \emph{Astronomische Nachrichten} \textbf{332}, 983 (2011)

\bibitem[{{Duez} et~al.(2010){Duez}, {Braithwaite}, \& {Mathis}}]{duez10b}
{Duez}, V., {Braithwaite}, J., {Mathis}, S., \emph{\apjl} \textbf{724}, 1, L34
  (2010)

\bibitem[{{Duez} \& {Mathis}(2010)}]{duez10a}
{Duez}, V., {Mathis}, S., \emph{\aap} \textbf{517}, A58 (2010)

\bibitem[{{Dufton} et~al.(2013)}]{dufton13}
{Dufton}, P.~L., et~al., \emph{\aap} \textbf{550}, A109 (2013)

\bibitem[{{Emeriau} \& {Mathis}(2015)}]{emeriau15}
{Emeriau}, C., {Mathis}, S., in G.~{Meynet}, C.~{Georgy}, J.~{Groh}, P.~{Stee}
  (eds.) New Windows on Massive Stars, \emph{IAU Symposium}, volume 307,
  373--374 (2015)

\bibitem[{{Emeriau} et~al.(2020){Emeriau}, {Mathis}, \& {Augustson}}]{mathis20}
{Emeriau}, C., {Mathis}, S., {Augustson}, K., \emph{\aap} \textbf{in prep.}
  (2020)

\bibitem[{{Featherstone} et~al.(2009){Featherstone}, {Browning}, {Brun}, \&
  {Toomre}}]{featherstone09}
{Featherstone}, N.~A., {Browning}, M.~K., {Brun}, A.~S., {Toomre}, J.,
  \emph{\apj} \textbf{705}, 1, 1000 (2009)

\bibitem[{{Fossati} et~al.(2015)}]{fossati15}
{Fossati}, L., et~al., \emph{\aap} \textbf{582}, A45 (2015)

\bibitem[{{Fossati} et~al.(2016)}]{fossati16}
{Fossati}, L., et~al., \emph{\aap} \textbf{592}, A84 (2016)

\bibitem[{{Fuller} et~al.(2015)}]{fuller15}
{Fuller}, J., et~al., \emph{Science} \textbf{350}, 6259, 423 (2015)

\bibitem[{{Garaud}(2018)}]{garaud18}
{Garaud}, P., \emph{Annual Review of Fluid Mechanics} \textbf{50}, 1, 275
  (2018)

\bibitem[{{Groh} et~al.(2013){Groh}, {Meynet}, {Georgy}, \&
  {Ekstr{\"o}m}}]{groh13}
{Groh}, J.~H., {Meynet}, G., {Georgy}, C., {Ekstr{\"o}m}, S., \emph{\aap}
  \textbf{558}, A131 (2013)

\bibitem[{{Huang} et~al.(2010){Huang}, {Gies}, \& {McSwain}}]{huang10}
{Huang}, W., {Gies}, D.~R., {McSwain}, M.~V., \emph{\apj} \textbf{722}, 1, 605
  (2010)

\bibitem[{{Jin} et~al.(2015){Jin}, {Zhu}, \& {L{\"u}}}]{jin15}
{Jin}, J., {Zhu}, C., {L{\"u}}, G., \emph{\pasj} \textbf{67}, 19 (2015)

\bibitem[{{Kaspi} \& {Beloborodov}(2017)}]{kaspi17}
{Kaspi}, V.~M., {Beloborodov}, A.~M., \emph{\araa} \textbf{55}, 1, 261 (2017)

\bibitem[{{Kochukhov} et~al.(2011){Kochukhov}, {Lundin}, {Romanyuk}, \&
  {Kudryavtsev}}]{kochukhov11}
{Kochukhov}, O., {Lundin}, A., {Romanyuk}, I., {Kudryavtsev}, D., \emph{\apj}
  \textbf{726}, 1, 24 (2011)

\bibitem[{{Kochukhov} et~al.(2013){Kochukhov}, {Mantere}, {Hackman}, \&
  {Ilyin}}]{kochukhov13}
{Kochukhov}, O., {Mantere}, M.~J., {Hackman}, T., {Ilyin}, I., \emph{\aap}
  \textbf{550}, A84 (2013)

\bibitem[{{Korre} et~al.(2019){Korre}, {Garaud}, \& {Brummell}}]{korre19}
{Korre}, L., {Garaud}, P., {Brummell}, N.~H., \emph{\mnras} \textbf{484}, 1,
  1220 (2019)

\bibitem[{{Krumholz}(2015)}]{krumholz15}
{Krumholz}, M.~R., {The Formation of Very Massive Stars}, \emph{Astrophysics
  and Space Science Library}, volume 412, 43 (2015)

\bibitem[{{Krumholz} \& {Federrath}(2019)}]{krumholz19}
{Krumholz}, M.~R., {Federrath}, C., \emph{Frontiers in Astronomy and Space
  Sciences} \textbf{6}, 7 (2019)

\bibitem[{{Langer}(2012)}]{langer12}
{Langer}, N., \emph{\araa} \textbf{50}, 107 (2012)

\bibitem[{{Ligni{\`e}res} et~al.(2014)}]{lignieres14}
{Ligni{\`e}res}, F., et~al., in P.~{Petit}, M.~{Jardine}, H.~C. {Spruit} (eds.)
  Magnetic Fields throughout Stellar Evolution, \emph{IAU Symposium}, volume
  302, 338--347 (2014)

\bibitem[{{MacDonald} \& {Petit}(2019)}]{macdonald19}
{MacDonald}, J., {Petit}, V., \emph{\mnras} \textbf{487}, 3, 3904 (2019)

\bibitem[{{Maeder}(2009)}]{maeder09}
{Maeder}, A., {Physics, Formation and Evolution of Rotating Stars} (2009)

\bibitem[{{Mathis}(2013)}]{mathis13}
{Mathis}, S., {Transport Processes in Stellar Interiors}, volume 865, 23 (2013)

\bibitem[{{Meynet} \& {Maeder}(2000)}]{meynet00}
{Meynet}, G., {Maeder}, A., \emph{\aap} \textbf{361}, 101 (2000)

\bibitem[{{Moravveji} et~al.(2016){Moravveji}, {Townsend}, {Aerts}, \&
  {Mathis}}]{moravveji16}
{Moravveji}, E., {Townsend}, R. H.~D., {Aerts}, C., {Mathis}, S., \emph{\apj}
  \textbf{823}, 2, 130 (2016)

\bibitem[{{M{\"o}sta} et~al.(2015)}]{mosta15}
{M{\"o}sta}, P., et~al., \emph{\nat} \textbf{528}, 7582, 376 (2015)

\bibitem[{Nagayama et~al.(2019)}]{nagayama19}
Nagayama, T., et~al., \emph{Phys. Rev. Lett.} \textbf{122}, 235001 (2019),
  \urlprefix\url{https://link.aps.org/doi/10.1103/PhysRevLett.122.235001}

\bibitem[{{Neiner} et~al.(2013){Neiner}, {Mathis}, {Saio}, \& {Lee}}]{neiner13}
{Neiner}, C., {Mathis}, S., {Saio}, H., {Lee}, U., in H.~{Shibahashi}, A.~E.
  {Lynas-Gray} (eds.) Progress in Physics of the Sun and Stars: A New Era in
  Helio- and Asteroseismology, \emph{Astronomical Society of the Pacific
  Conference Series}, volume 479, 319 (2013)

\bibitem[{{Neiner} et~al.(2012)}]{neiner12}
{Neiner}, C., et~al., \emph{\aap} \textbf{546}, A47 (2012)

\bibitem[{{Neiner} et~al.(2015)}]{neiner15}
{Neiner}, C., et~al., in K.~N. {Nagendra}, S.~{Bagnulo}, R.~{Centeno},
  M.~{Jes{\'u}s Mart{\'\i}nez Gonz{\'a}lez} (eds.) Polarimetry, \emph{IAU
  Symposium}, volume 305, 61--66 (2015)

\bibitem[{{Nomoto} et~al.(2013){Nomoto}, {Kobayashi}, \& {Tominaga}}]{notomo13}
{Nomoto}, K., {Kobayashi}, C., {Tominaga}, N., \emph{\araa} \textbf{51}, 1, 457
  (2013)

\bibitem[{{Ogilvie}(2014)}]{ogilvie14}
{Ogilvie}, G.~I., \emph{\araa} \textbf{52}, 171 (2014)

\bibitem[{{Pedersen} et~al.(2018){Pedersen}, {Aerts}, {P{\'a}pics}, \&
  {Rogers}}]{pedersen18}
{Pedersen}, M.~G., {Aerts}, C., {P{\'a}pics}, P.~I., {Rogers}, T.~M.,
  \emph{\aap} \textbf{614}, A128 (2018)

\bibitem[{{Prat} et~al.(2019)}]{prat19}
{Prat}, V., et~al., \emph{\aap} \textbf{627}, A64 (2019)

\bibitem[{{Pratt} et~al.(2017)}]{pratt17}
{Pratt}, J., et~al., \emph{\aap} \textbf{604}, A125 (2017)

\bibitem[{{Raghavan} et~al.(2010)}]{raghavan10}
{Raghavan}, D., et~al., \emph{\apjs} \textbf{190}, 1, 1 (2010)

\bibitem[{{Ram{\'\i}rez-Agudelo} et~al.(2013)}]{ramirez13}
{Ram{\'\i}rez-Agudelo}, O.~H., et~al., \emph{\aap} \textbf{560}, A29 (2013)

\bibitem[{{Romanova} \& {Owocki}(2015)}]{romanova15}
{Romanova}, M.~M., {Owocki}, S.~P., \emph{\ssr} \textbf{191}, 1-4, 339 (2015)

\bibitem[{{Sengupta} \& {Garaud}(2018)}]{sengupta18}
{Sengupta}, S., {Garaud}, P., \emph{\apj} \textbf{862}, 2, 136 (2018)

\bibitem[{{Smith}(2014)}]{smith14}
{Smith}, N., \emph{\araa} \textbf{52}, 487 (2014)

\bibitem[{{Stello} et~al.(2016)}]{stello16}
{Stello}, D., et~al., \emph{\nat} \textbf{529}, 7586, 364 (2016)

\bibitem[{{Sundqvist} et~al.(2013)}]{sundqvist13}
{Sundqvist}, J.~O., et~al., \emph{\mnras} \textbf{433}, 3, 2497 (2013)

\bibitem[{{Tan} et~al.(2014)}]{tan14}
{Tan}, J.~C., et~al., in H.~{Beuther}, R.~S. {Klessen}, C.~P. {Dullemond},
  T.~{Henning} (eds.) Protostars and Planets VI, 149 (2014)

\bibitem[{{Viallet} et~al.(2013){Viallet}, {Meakin}, {Arnett}, \&
  {Moc{\'a}k}}]{viallet13}
{Viallet}, M., {Meakin}, C., {Arnett}, D., {Moc{\'a}k}, M., \emph{\apj}
  \textbf{769}, 1 (2013)

\bibitem[{{Vidal} et~al.(2018){Vidal}, {C{\'e}bron}, {Schaeffer}, \&
  {Hollerbach}}]{vidal18}
{Vidal}, J., {C{\'e}bron}, D., {Schaeffer}, N., {Hollerbach}, R., \emph{\mnras}
  \textbf{475}, 4, 4579 (2018)

\bibitem[{{Vidal} et~al.(2019){Vidal}, {C{\'e}bron}, {Ud-Doula}, \&
  {Alecian}}]{vidal19}
{Vidal}, J., {C{\'e}bron}, D., {Ud-Doula}, A., {Alecian}, E., \emph{arXiv
  e-prints} arXiv:1902.10599 (2019)

\bibitem[{{Wade} et~al.(2014)}]{wade14}
{Wade}, G.~A., et~al., in P.~{Petit}, M.~{Jardine}, H.~C. {Spruit} (eds.)
  Magnetic Fields throughout Stellar Evolution, \emph{IAU Symposium}, volume
  302, 265--269 (2014)

\bibitem[{{Woosley} et~al.(2007){Woosley}, {Blinnikov}, \& {Heger}}]{woosley07}
{Woosley}, S.~E., {Blinnikov}, S., {Heger}, A., \emph{\nat} \textbf{450}, 7168,
  390 (2007)

\end{thebibliography}

\end{document}